\documentclass[%
 reprint,
 amsmath,amssymb,
 aps,
]{revtex4-2}

\usepackage{graphicx}
\usepackage{dcolumn}
\usepackage{bm}
\usepackage[utf8]{inputenc}
\usepackage{CJKutf8}
\usepackage{float}

\begin{document}
\title{Formation and mechanics of fire ant rafts as an active self-healing membrane}

\title{Formation and mechanics of fire ant rafts as an active self-healing membrane}
\author{Chung-Hao Chen$^1$, Ting-Heng Hsieh$^1$, Hong-Yue Huang$^1$, Yu-Chuan Cheng$^{1,2}$, and Tzay-Ming Hong$^{1}$\thanks{ming@phys.nthu.edu.tw} }
\thanks{ming@phys.nthu.edu.tw}
\affiliation{$^1$ Department of Physics, National Tsing Hua University, Hsinchu, Taiwan 30013, Republic of China\\
$^2$ Department of Physics, University of California at Santa Barbara, Santa Barbara, California 93106, USA}
\date{\today}

\begin{abstract}

The unique ability of fire ants to form a raft to survive flooding rain has enchanted biologists as well as researchers in other disciplines. It has been established during the last decade that an aggregation of fire ants exhibits viscoelasticity with respect to external compression and shearing among numerous unusual mechanical properties. In addition to clarifying that Cheerios effect is neither sufficient nor essential for the ant raft, we perform the force-displacement and creep experiments on the ant raft and concentrate on unearthing properties that derive from the unique combination of self-healing and activeness of its constituent. Varying the pull speed results in distinct mechanical responses and fracture patterns,  characteristic of ductile and brittle material. By image processing, we count the number of ants that actively participate in the stress-strain relation and determine their orientation to map out the force chain. The latter information reveals that the  pull force  expedites the alignment of fire ants, in analogy to the effect of electric field on liquid crystal polymers. 
In addition, the raft can be tailored not to transversely deform in response to the axial strain.
Without resorting to specific geometry structures, this property of zero Poisson's ratio is enabled by the active recruitment of ants from the top to bottom layer to keep the raft from disintegrating. Furthermore,  effective Young's modulus can also be customized and is proportion to either the raft length or its inverse, depending on whether the raft is in the elastic or plastic region.
\end{abstract}

\maketitle

\section{introduction}

Examples of collective behavior, shaped by hundred million years of evolution  to adapt to the changing environments, are ubiquitous in the animal kingdom, such as the bird flocking  \cite{pearce2014role, bialek2012statistical}
, cell motion  \cite{szabo2010collective} and bacterial growth \cite{mattingly2022collective}. One recent development in bionics is to synthesize materials that exhibit similar emergent phenomena, like Janus particles  \cite{walther2013janus, elgeti2015physics}, robots with local sensing  \cite{werfel2014designing} and microtubule-based liquid crystal  \cite{sanchez2012spontaneous}.
This inclination to socialize and collaborate  makes sense  particularly for the tiny ants to bring back a large food or fight off an invading enemy. What's amazing is that some species of ant  develop the skill to aggregate and form a thread to cross the gap of a hiatus or to reach a higher ground.
In the last eighty years or so, fire ants have attracted much attention   not only because their global invasion causes billions of agricultural loss, medical treatments, and damage to properties annually in the US  \cite{ascunce2011global}, but for academic reasons. The latter has to do with their unique ability to form rafts in order to float and survive the flooding. This has captivated the imagination of material engineers and physicists who already succeeded at characterizing their viscoelastic and other distinct mechanical properties in the form of a raft or bulk  \cite{tennenbaum2016mechanics, tennenbaum2017activity, tennenbaum2020activity}. 

One explanation for how ants make a raft was proposed by Ko {\it et al.}  \cite{PhysRevFluids.7.090501}  that it is due to Cheerios effect\cite{{vella2005cheerios}} , i.e.,  the surface tension. They numerically simulated the ants to repel each other due to panic, and it was only after the water was taken into account that their aggregation became possible. However, a large crowd, roughly exceeding seven, was found to be necessary to stabilize the raft  \cite{PhysRevFluids.7.090501}. The clumping of cereals that float in our  bowl of milk at breakfast is so common that their explanation seems intuitive at first glance. But why didn't this ability to form a raft on water also exist in other species of ants? To solve this dilemma, we arranged to put live/dead fire ants and a different species of ant on a shaker to test if the crisis can prompt the action of aggregation, and record how fast the ants assemble into a raft and how long it takes for the ants to dissolve after the raft is allocated to a solid surface.

Continuing the nice work by Tennenbaum {\it et al.} \cite{tennenbaum2016mechanics} who studied the viscoelastic and fluid-like properties by oscillating and rotating 3D fire ant aggregates, we pay attention to  the 2D ant raft formed on water and perform uniaxial tensile test and linear creep experiment  to search for new characteristics driven by active and self-healing nature. They include measuring the responses at different pull speeds, such as the resistance force, crack modes and Young's modulus during relaxation, and then comparing them with the common tissue and bubble raft  \cite{lauridsen2004velocity,wang2006impact,bowick2008bubble,twardos2005comparison,ritacco2007lifetime,dennin2004statistics,lauridsen2002shear,wang2006bubble,kuo2013scaling} to accentuate the effect when the constituents can not only actively adjust their position and role in response to external forces, but also possess the healing power. This information is valuable since, although literature is abound for either active or self-healing material \cite{activematerial_1,activematerial,selfhealingmaterial,selfhealingmaterial_2,selfhealingmaterial_3}, their combination is rare.

Bearing in mind that material failure is usually classified into brittle\cite{persson2005crack, fleck2011brittle, mulla2018crack} and ductile \cite{kayagaki2021ninj1,fineberg1999instability, mlot2011fire,marder1996energetic} where the former has nearly no plastic deformation and breaks down like an avalanche at the macroscopic scale. In contrast, the fracture processes in a ductile material are more complex that holes appear randomly on the film and eventually cause fracture  upon extension. Similar comparison has been made to the fire ant raft by concentrating on the viscoelasticity and activeness. Besides elaborating on these macroscopic properties, we zoom in each ant and analyze how the force chain develops and propagates with the pull force and relaxation time. The structure of force chain is hard to determine, and has only been studied in granular material  \cite{majmudar2005contact, daniels2017photoelastic} because its constituents are constantly in the jam state without external vibrations. The  Janssen effect, observed for a fire-ant column by measuring its apparent mass \cite{anderson2022janssen}, indirectly confirmed the existence of force chain, but failed to provide any information on its dynamics.

\section{Experimental Setup}

\begin{figure}
\includegraphics[scale=0.6]{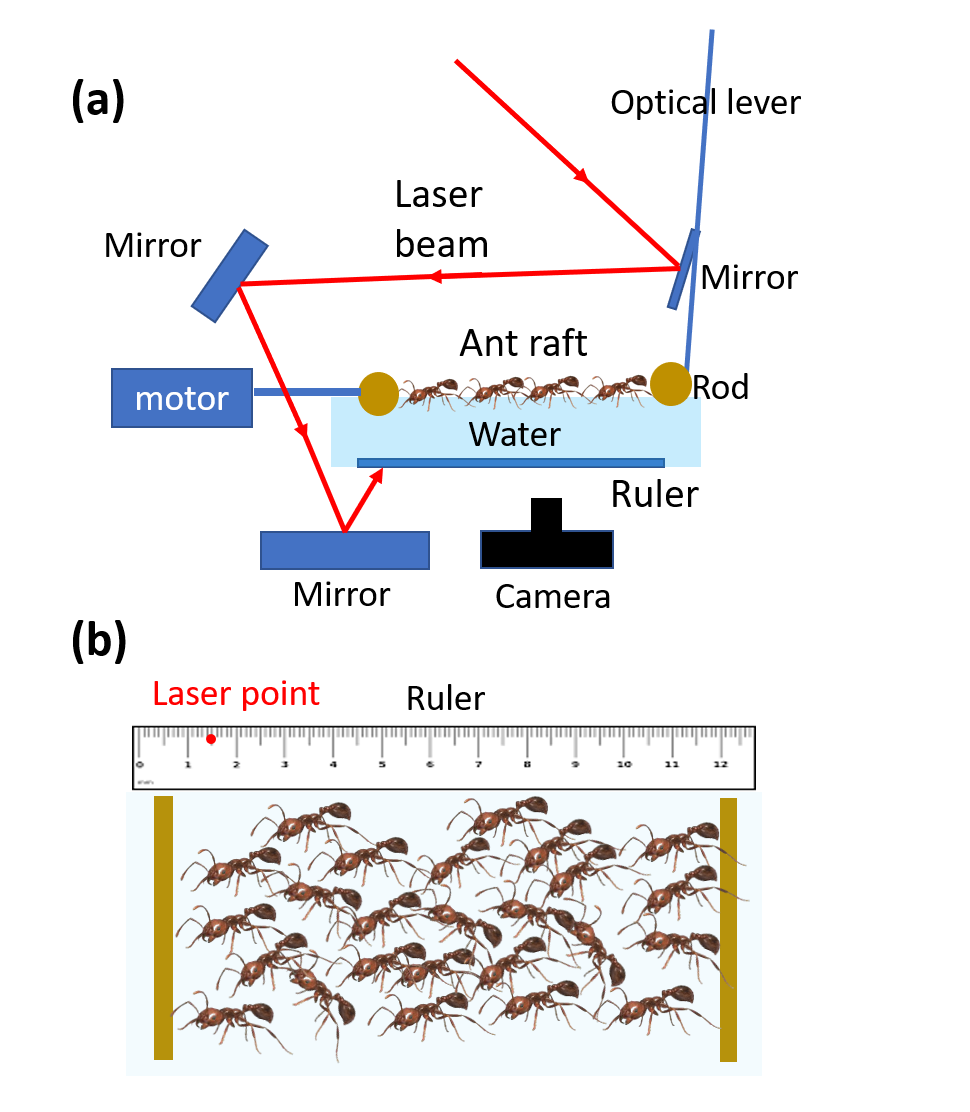}
\caption{(a) Schematic experimental setup
for measuring the mechanical and creep properties of fire ant raft.  (b) The displacement of laser point on the ruler allows us to determine the bending angle of optical lever and hence the resistance force of ants from this bottom view of (a). }
\label{setup}
\end{figure}
\begin{figure}
    \centering
    \includegraphics[scale=0.4]{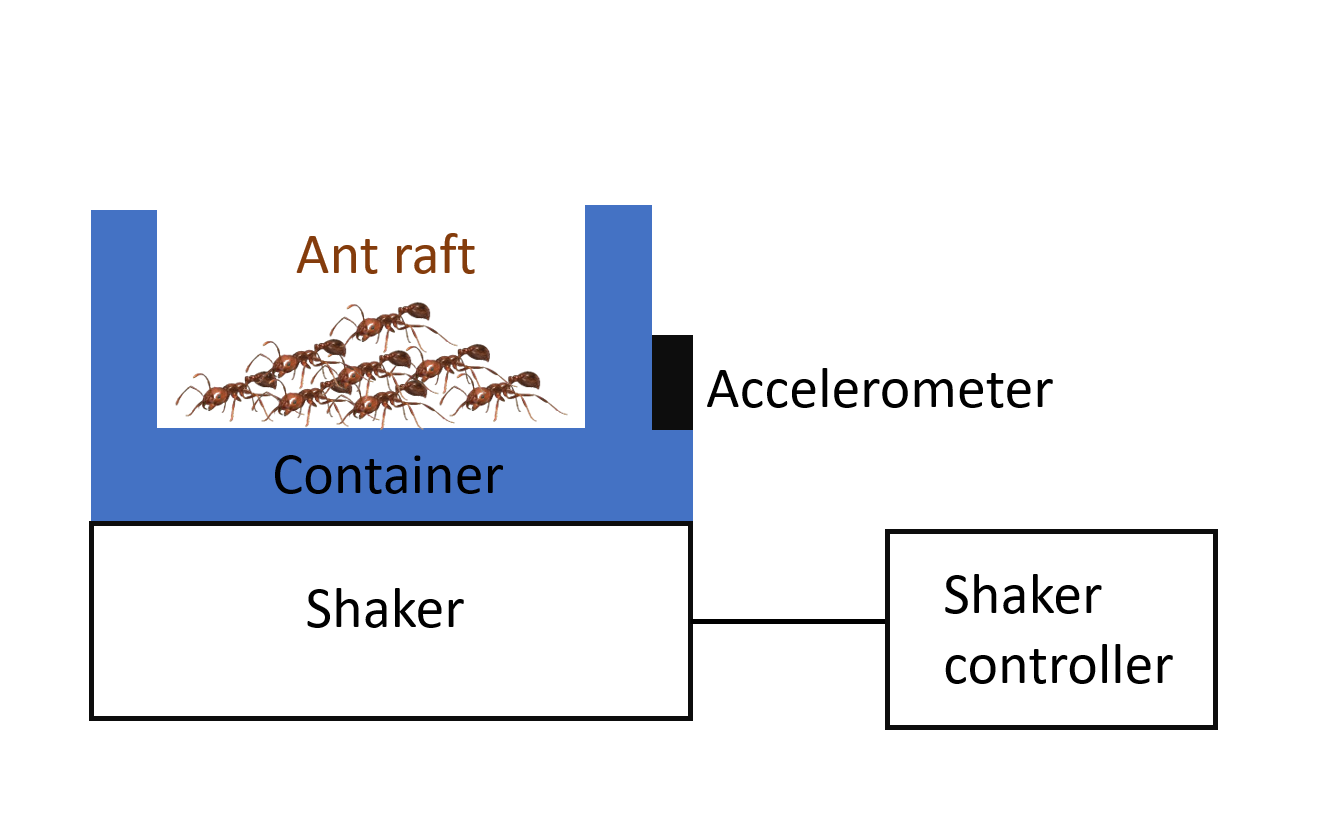}
    \caption{A shaker is employed to test whether flooding is the only means to induce the action of aggregation for fire ants. Note that ants never touch the wall when jiggled horizontally. 
    The container is coated with teflon (Chemours DISP 30) to  avoid fire ants from gripping and escaping.   }
    \label{shaker}
\end{figure}

Our samples consist of fire ant ({\it Solenopsis invicta}) and a common species ({\it Monomorium chinense}) in Taiwan, both of which are approximately $5$ mm in length.
 We adopted the methods  by Chen  \cite{chen2007advancement} to retrieve ants from the field and separate them from the soil. Once separated and  placed into a plaster nest, ants are fed with sugar, cricket and water. To measure the resistance force $F$ of ants as we pull the raft, a home-made force sensor is designed with an optical lever \cite{guo2013direct}, as shown in Fig. \ref{setup}.
 For the ants to cling onto, two wooden rods of diameter = 1 cm are arranged on both sides of the raft. While one rod is fixed to the lever, the other is pulled by a motor at a constant speed $v$. The optical lever consists of a  mirror and a thin glass fiber of length = 50 cm and diameter = 0.1 cm.  The morphology of ant raft is video-recorded simultaneously with the resistance force. In addition, the number of ants at the bottom layer, which are mostly static and believed to compose the raft and contribute to counter $F$, and their orientation are determined by image processing.
To test whether aggregation can be induced by means other than flooding, we bring into action the shaker in Fig. \ref{shaker} with the frequency range 35$\sim$55 Hz and root-mean-square acceleration 1$\sim$3 g where g = 9.8 $\rm m/s^2$. It is enabled by an accelerometer and switchable between vertical and horizontal modes.

\begin{table*}
\begin{center}
\setlength{\tabcolsep}{0.1mm}{
\begin{tabular}{clclclcl} 
\hline\hline
                    & \quad\quad\ \ \ \ \ \ \ \ \ \ \ \ \ \ \ \ \ \  on water                                                 & \ vertical shaking                                                     & \ \ \ \ \ \ \ \ \ \ \ horizontal shaking                                           \\ 
\hline 
 fire ants           & \begin{tabular}[c]{@{}l@{}}\\ \includegraphics[height=0.7 in]{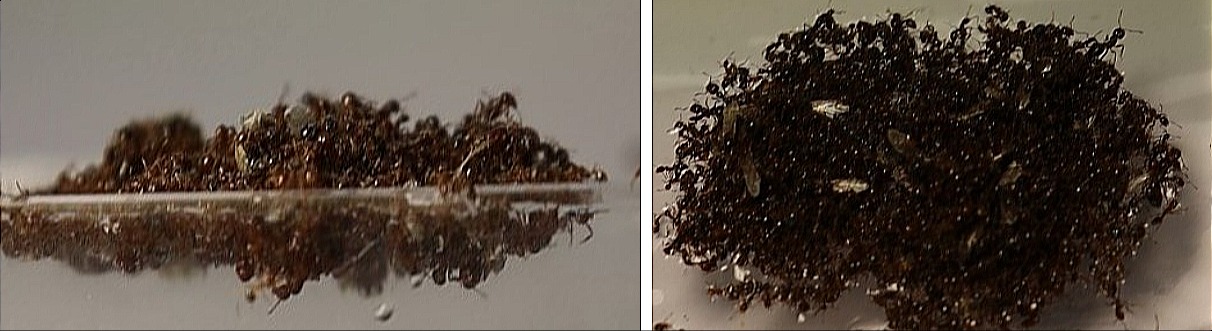}  \end{tabular} &\ \ \begin{tabular}[c]{@{}l@{}}\\ \includegraphics[height=0.7in]{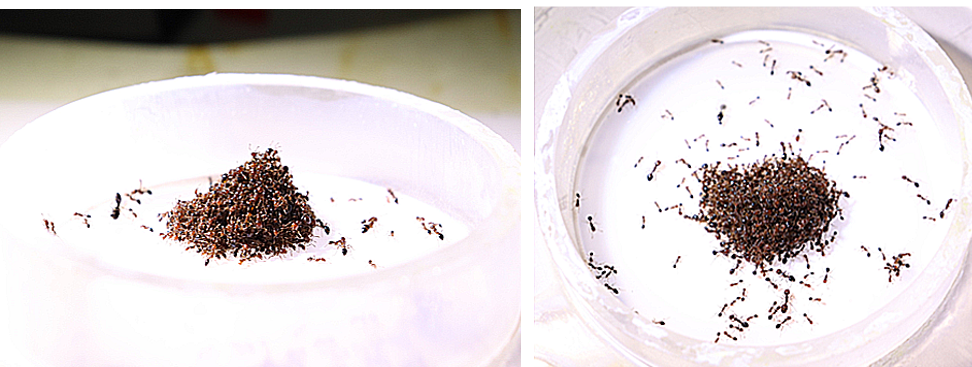}  \end{tabular} &\ \ \  \begin{tabular}[c]{@{}l@{}}\\ \includegraphics[height=0.7in]{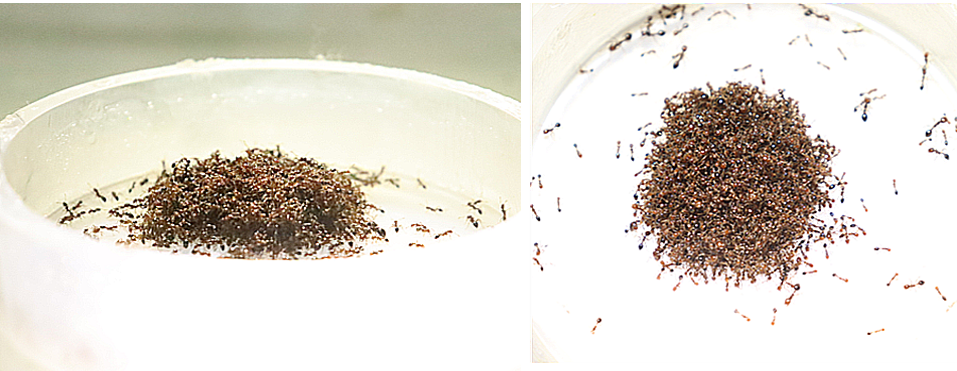}  \end{tabular}  \\
local ants & \begin{tabular}[c]{@{}l@{}}\\ \includegraphics[height=0.7 in]{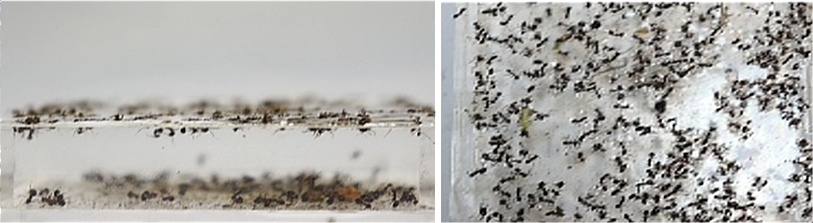}  \end{tabular} &\ \ \begin{tabular}[c]{@{}l@{}}\\ \includegraphics[height=0.7in]{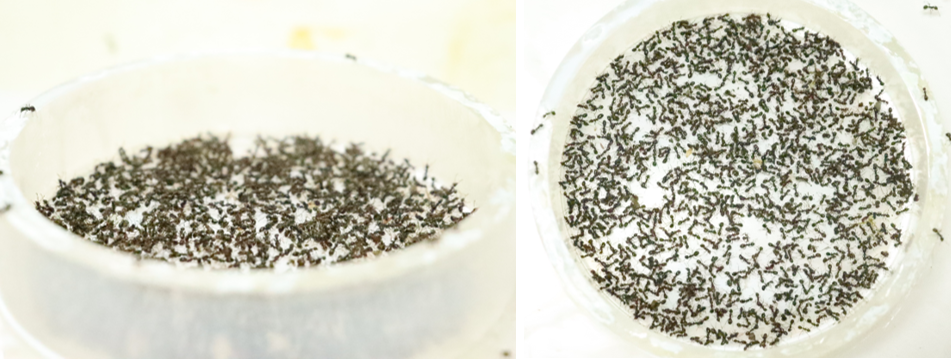}  \end{tabular} &\ \ \   \begin{tabular}[c]{@{}l@{}}\\ \includegraphics[height=0.7in]{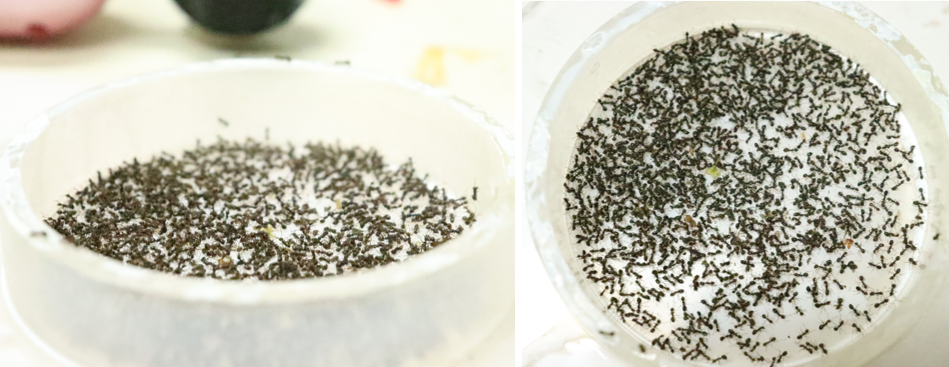}  \end{tabular}  \\
dead ants           & \begin{tabular}[c]{@{}l@{}}\\ \includegraphics[height=0.7 in]{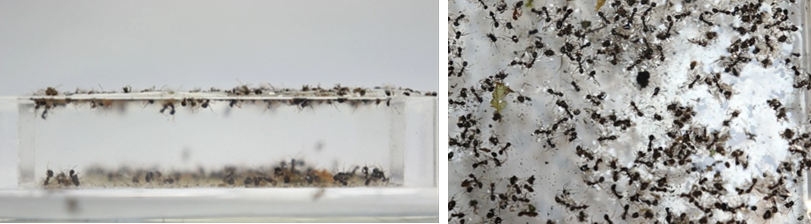}  \end{tabular} &\ \ \begin{tabular}[c]{@{}l@{}}\\ \includegraphics[height=0.7in]{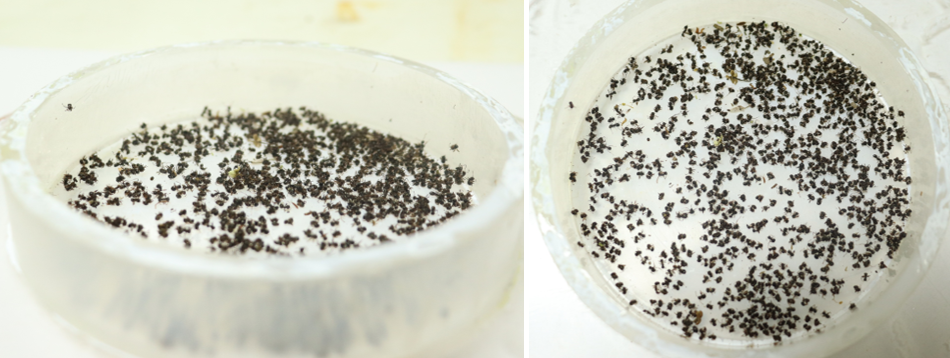}  \end{tabular} &\ \ \   \begin{tabular}[c]{@{}l@{}}\\ \includegraphics[height=0.7in]{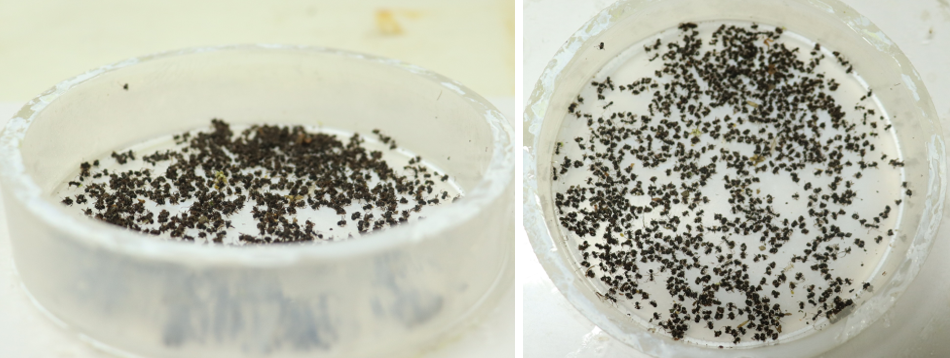}  \end{tabular}  \\
\hline\hline
\end{tabular}}
\end{center}
\caption{Side and top views of fire ants, local ants and dead fire ants under different environments to facilitate aggregation. }
\label{ant_shape}
\end{table*}

\begin{table*}
\centering
\begin{tabular}{clclclclclclclclc} 
\hline\hline
           &          &                  &                     &  & \multicolumn{3}{l}{}                               \\
           &          & \quad \quad \quad \quad   aggregation time      &                     &  & \multicolumn{3}{l}{\quad \quad \quad \quad \quad  relaxation time}                     \\
           &          &                  &                     &  & \multicolumn{3}{l}{}                               \\ 
\cline{2-4}\cline{6-8}
           &          &                  &                     &  & \multicolumn{3}{l}{}                               \\
           & on water & vertical shaking$ ^\star$ & horizontal shaking$ ^\star$ &  & on water & vertical shaking & horizontal shaking  \\
           &          &                  &                     &  &          &                  &                      \\ 
\hline
           &          &                  &                     &  &          &                  &                      \\
fire ants           & \begin{tabular}[c]{@{}l@{}}\quad 30 sec\end{tabular} & \begin{tabular}[c]{@{}l@{}}5 sec \end{tabular} & \begin{tabular}[c]{@{}l@{}}\quad \quad 3 sec\end{tabular} &  & \ 60 sec  & 10 sec       & \quad \quad 30 sec             \\
           &          &                  &                     &  &          &                  &                      \\
local ants & \begin{tabular}[c]{@{}l@{}}\quad \ \ \ X\end{tabular} & \begin{tabular}[c]{@{}l@{}}\ \ \ X\end{tabular} & \begin{tabular}[c]{@{}l@{}}\ \ \ \ \ \ \ \ \ X\end{tabular} &  & \ \ \ \  X  & \ \ \ X          & \quad \quad \ \ \ \ \ X              \\
           &          &                  &                     &  &          &                  &                      \\
dead fire ants  & \begin{tabular}[c]{@{}l@{}}\quad \ \ \ X\end{tabular}& \quad X                & \quad \ \ \  1 sec                  &  & \begin{tabular}[c]{@{}l@{}}\quad \   X\end{tabular}        & \ \ \  X                & \ \ \ \ \ \ \ \ \ \ \ $\infty $                  \\
           &          &                  &                     &  &          &                  &                      \\
\hline\hline
\end{tabular}
\caption{The characteristic aggregation and relaxation time for ant rafts formed on water and by vertical and horizontal shaking. At the beginning, 500 ants  are randomly scattered, and the aggregation  time is defined as the time it takes for half of the ants to cling together. The relaxation time measures how long it takes for the raft to disintegrate after being relocated to a stationary solid surface.  The cross mark means that the ants fail to aggregate.\\ $^\star$amplitude $ = $ 0.25 mm and frequency $ = $ 50 Hz. }
\label{time}
\end{table*}

\section{Experimental results}
\subsection{Raft formed by different means}
\begin{figure*}
    \centering
    \setlength{\leftskip}{-12pt}
    \includegraphics[scale=0.39]{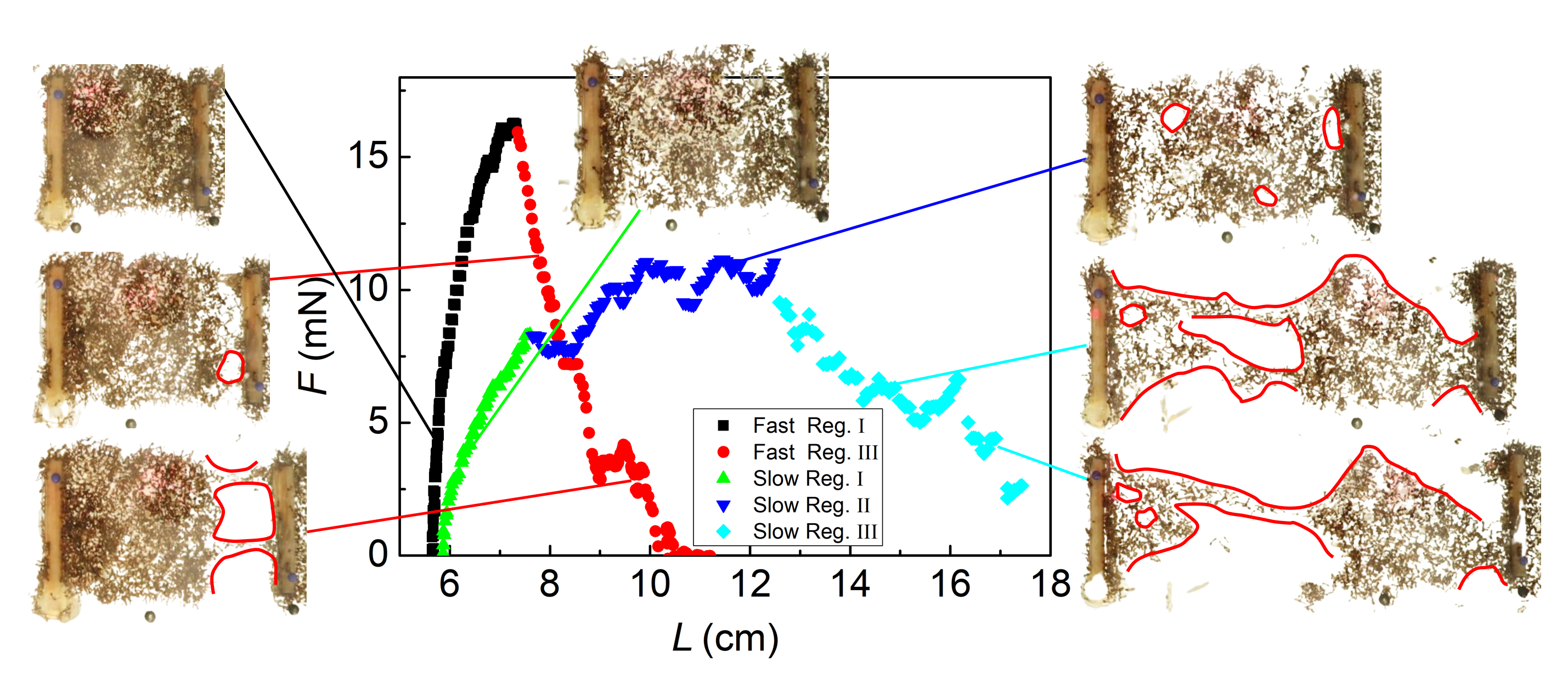}
    \caption{ The pull force versus length of ant raft under  fast (1.7 mm/s) and slow (0.25 mm/s)  pull speeds. Representative snapshots for each regions are shown. Red curves are used to highlight the cracks developed in the interior and at the edge of the ant raft. In contrast to being stationary in region II, these cracks expand irreversibly upon entering region III. Note that region II is missing for high pull speed.}
\label{pulling}
\end{figure*}

By tuning the frequency and amplitude of shaker in Fig. \ref{setup}(b), we can simulate a zero-gravity environment. As we suspected, local ants and dead fire ants remain scattered on water, while live fire ants do  aggregate under both vertical and horizontal shaking, as shown in Table \ref{ant_shape}. This proves that Cheerios effect is neither sufficient nor necessary for the raft formation.
There is a slight difference in morphology, though. The vertical shaking generates a ball-like structure,  conspicuously different from the double-deck raft formed on water. This is in contrast to the pancake-like aggregate triggered by horizontal shaking. No raft formation is possible for local ants in all three circumstances, likely due to the repulsion from their vigorous waving of legs. 

In addition to  morphology, ant aggregates formed by different means can also be distinguished by the time it takes for the ants to accumulate or disperse after the cluster is relocated to a safe and firm ground, as compiled in Table \ref{time}.
 Unlike water on which ants can float and survive for dozens of minutes, shaking is an imminent danger that  calls for an immediate action and thus explains the shorter aggregation time. But why is the relaxation time more than twice as the aggregation time? We suspect  it is constrained by the time it takes for the pheromone to disperse so that the intoxicated ants can finally sober up and disseminate. In the mean time, it goes without saying that the dead  ant raft never relaxes.

\subsection{Force-displacement experiment}

Had the raft been formed by Cheerios effect, one would have expected similar behavior under a linear extension to the membrane formed by plastic balls in Ref.\cite{to2022rifts}. But they turn out to exhibit some very different mechanical behavior, except for the sharing of two distinct crack patterns determined by the magnitude of pull speed $v$. In the case of ant raft, the latter property is ascribed to the agility of ants which sets an intrinsic speed for response and recovery. 
As in most ductile material like gold leaf and clay, the stress-strain curve of fire ant raft  in Fig. \ref{pulling} can be separated into elastic, plastic and rupture regions for a low pull speed $v=0.25$ mm/s, denoted respectively by Reg. I, II and III. Region II is characterized by the appearance of stationary cracks, first in the interior and then at the edge of the ant raft. These cracks start to expand irreparably upon entering region III  before eventually breaking into two pieces at some critical raft length that decreases as $v$ is raised. 
The plastic region II is lacking as we switch to a high speed $v$ = 1.7 mm/s, reminiscent of brittle material.
Again similar to normal substance, the maximum resistance force is enhanced when pulled faster. 
In contrast to other factors that causes brittleness in a ductile material, such as temperature, load, strength and stress concentration, the sensitive dependence on the stress rate is more like the viscoelastic material, e.g., amorphous polymers and non-Newtonian liquids such as oobleck. 
\begin{figure*}
    \centering          
    \setlength{\leftskip}{-12pt}
   \includegraphics[scale = 0.63]{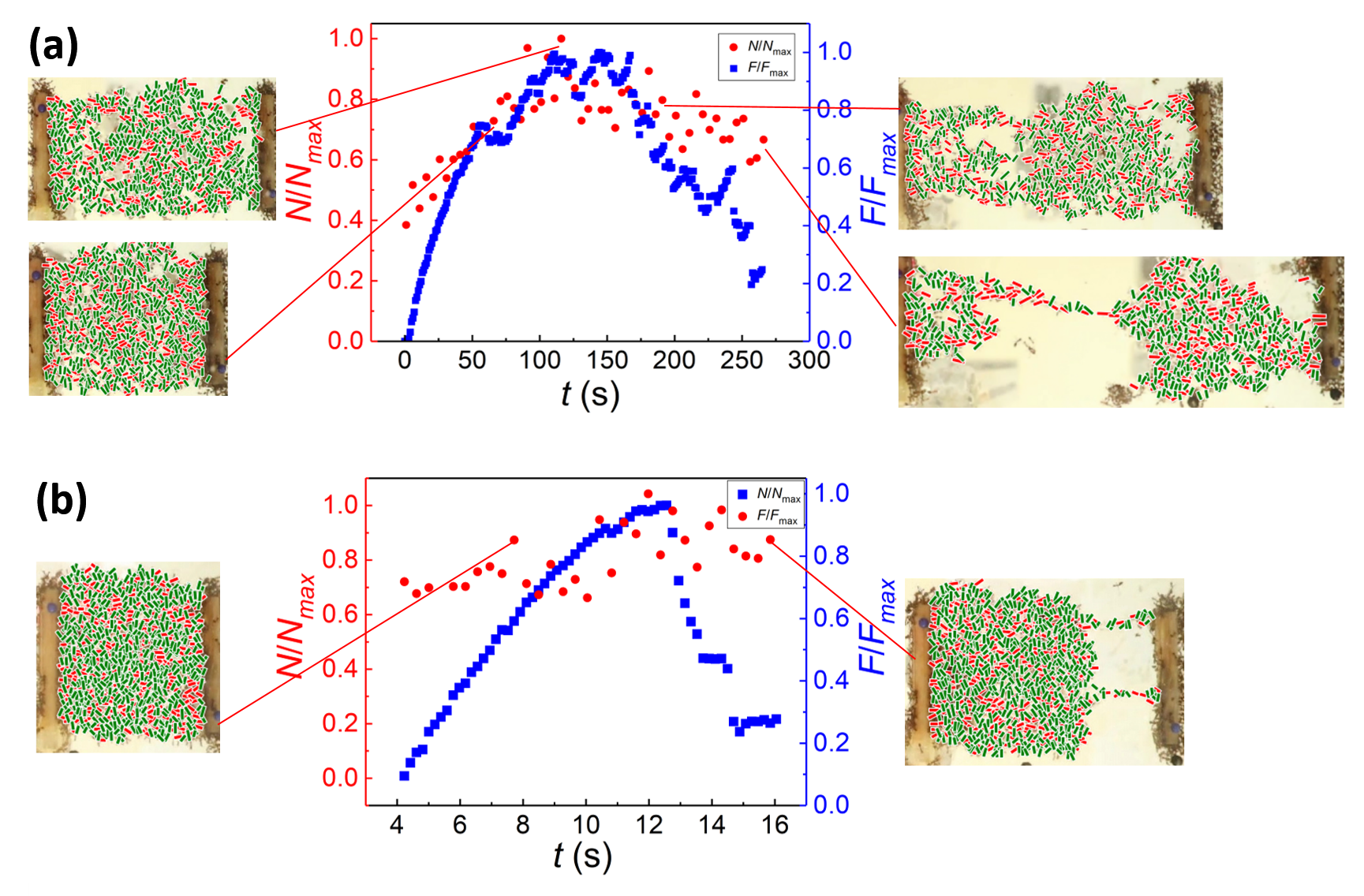} 
   \caption{In order to capture how the number of aligned ants $N$ is correlated with the pull force $F$, they are  plotted in double $y$-axis graphs as a function of time. Both quantities are rescaled by their maximum value to facilitate comparison.  (a) and (b) are respectively  for low and high pull speed. Alignment is defined by an  deviation less than $20^{\circ}$ between the ant orientation and the pull direction. These ants are marked by red lines in the photos,  otherwise in green. }
    \label{slow alpha}
\end{figure*}

\begin{figure}
    \centering
    \setlength{\leftskip}{4pt}
    \includegraphics[scale = 0.75]{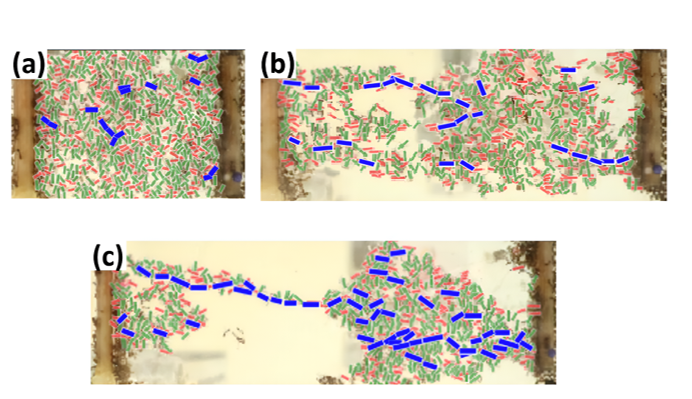}
    \caption{Visualized by connecting the blue lines as defined in the text, force chains gradually develop from regions I and II in (a, b)  to III in (c) for low pull speed.  }
    \label{force chain}
\end{figure}

In addition to mechanical and morphological  descriptions, we also paid attention to the detailed alignment of ants by taking photos of raft from below.
The ants on the top layer are mobile and do not actively contribute to $F$ and the raft formation, but are important as a reserve to be called up in times of need. Aided by image processing and comparing successive photos in time, we single them out when counting the active-duty ants at the bottom layer. Among them, those with orientation within $20^{\circ}$ from the pull direction are tallied as number $N$ and marked by red lines, otherwise in green. The degree of alignment is shown in Fig. \ref{slow alpha} to correlate positively with the pull force, in analogy to the liquid crystal under an external electric field. The correlation is stronger in Fig. \ref{slow alpha}(a)  than (b) presumably due to the ample time allowed by the low pull speed for ants to readjust their position to counter the applied force.

How to extract information about the structure of force chains from Fig. \ref{slow alpha}?
Although the red lines already exist when the orientation of ants is supposed to be random at $F$=0 from Fig. \ref{slow alpha}, we expect them to appear arbitrarily in space. Thus, by paying attention to only those that are nearest neighbors, we can eliminate the randomness and identify  the ants that actually contribute to the force chains. These pairs of red lines from Fig. \ref{slow alpha}(a) are replaced by blue ones. The fact that  blue lines are patchy in  Fig. \ref{force chain}(a) for region I reflects the truth, and intrinsic challenge at doing such a map, that not all ants on the force chain are aligned to the force direction. Indisputably, the structure of force chain becomes clearer in Fig. \ref{force chain}(b, c) for regions II and III when the number of aligned ants increases.
 
\subsection{Creep Experiment}
\begin{figure}[ht]
    \centering
    \setlength{\leftskip}{-10pt}
    \includegraphics[scale= 0.34]{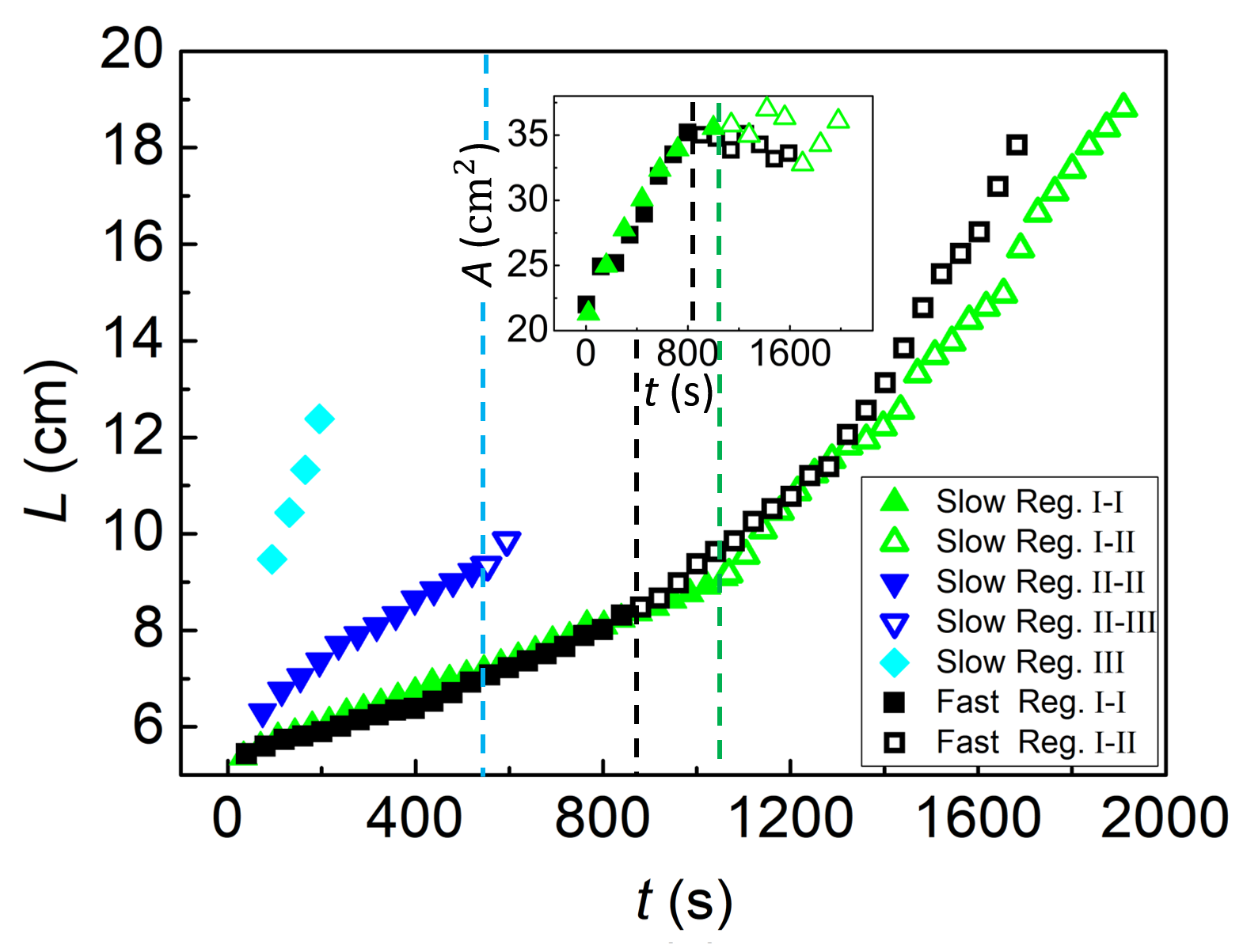}
    
    \caption{ The creep experiment 1: Length $L$ of ant raft relaxes under a constant $F$ = 2.46, 13.43 and 8.93 mN for Slow Reg. I, II and III, and 3.4 mN for Fast Reg. I. The blue dashed line denotes the transition for Slow Reg. II-III data, i.e., for  a slow pull speed and from region II to III. Likewise, yellow is for Fast Reg. I-III, and red is for Slow Reg. I-II.  Inset shows the change of morphology for yellow and red lines is accompanied by an approximate saturation of raft area, $A$.
    }
    \label{creep}
\end{figure}

\begin{figure}
    \centering
    \setlength{\leftskip}{-15pt}
    \includegraphics[scale = 0.3]{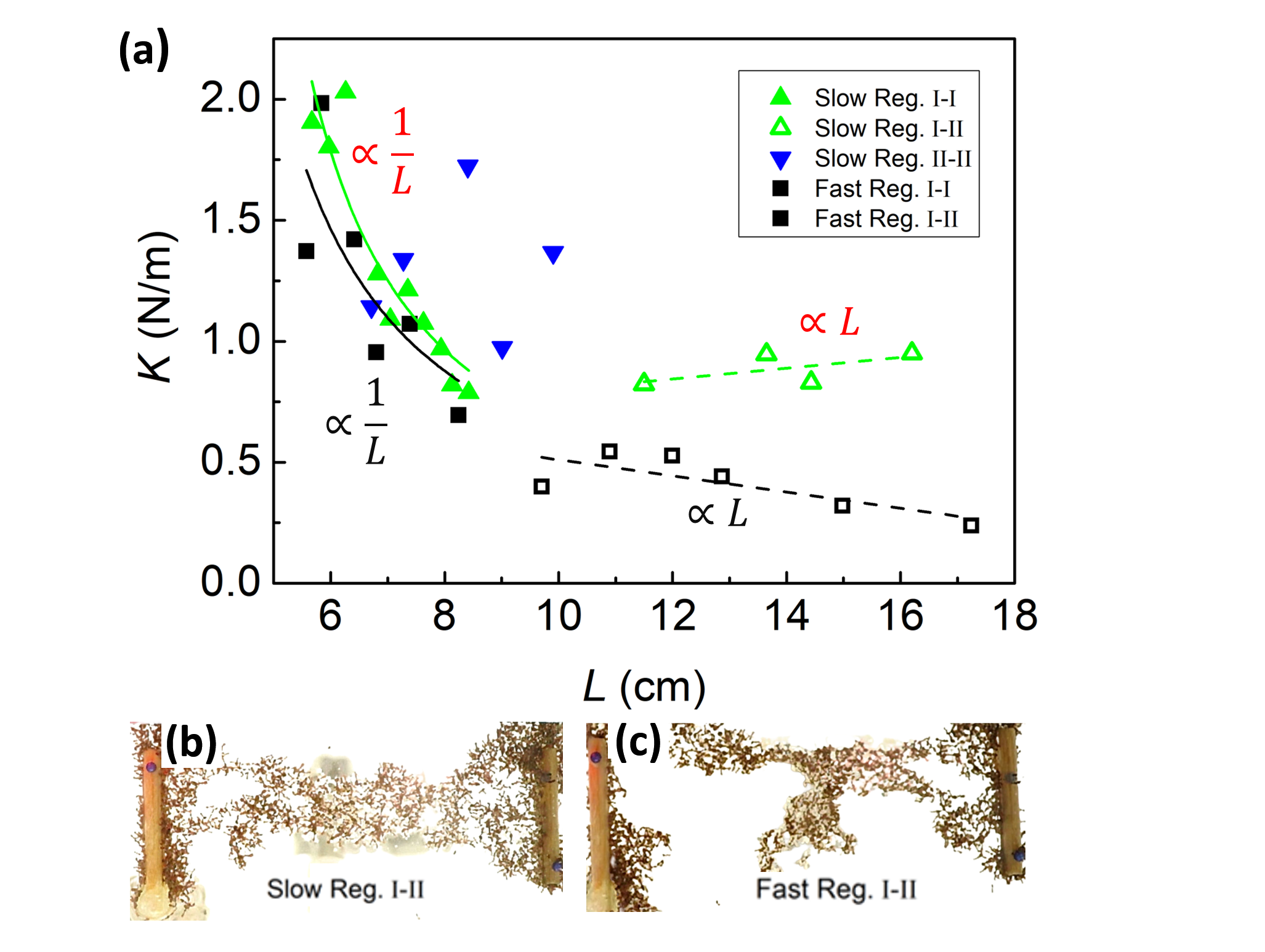}
    \caption{The creep experiment 2: (a) Effective spring constant $K$  is plotted against  $L$ in Fig. \ref{creep}. Except for  Slow Reg. II-II, the other four data are fitted respectively by solid ($1/L$) and dashed ($L$) green lines for slow speed, and the same  in black  for fast speed. Although roughly observing the increasing trend as Slow Reg. I-II, the data of Slow Reg. II-II fluctuate wildly due to the fact that we are forced to adopt a smaller $\Delta F$ to avoid hastening the raft into region III. This increases the uncertainty in $\Delta L$. Snapshots of Slow and Fast Reg. I-II rafts are shown in (b) and (c). Note that the neck in (c) is not only
narrower than that in (b), but its width also shrinks faster. }
    \label{K vs L}
\end{figure}

The magic of fire ant raft, treated as a novel membrane, lies in the fact that its ingredient is both active and capable of self-healing. In order to elucidate the effect of these characteristics on the material properties, we perform the creep experiment.  As shown in Fig. \ref{creep}, region-I raft can relax to region II and III over time, even for samples prepared by fast pull speed. Although no crack appears throughout region I, $L$ is found to increase, while the width and height of the raft remain roughly constant. This implies  Poisson's ratio equals zero in this region. Such a property is made possible by the active movement of ants from the top to the bottom layer to help reinforce and keep  the raft intact. As the raft runs out of  ants on the top layer, two things happen  as it enters region II: $A$ begins to saturate and cracks start to develop. The latter weakens the structure of ant raft and gives rise to a steeper slope in Fig. \ref{creep}. Note that this is opposite to the behavior of most ductile material where plastic deformation, movements and  generation of dislocation to be specific, strengthens the metal or polymer via work  hardening.
Since region-III raft by a fast pull speed is very unstable, we only experiment on the sample prepared by a slow one. By definition, the existing cracks start to expand irreversibly in this region, which thus renders an even larger $dL/dt$ than that of region II in Fig. \ref{creep}.

Equivalent to Young's modulus, an effective spring constant is defined as $K\equiv\Delta F/\Delta L$ where  $\Delta L$ is the slight displacement to achieve a destined increment of resistance force $\Delta F$ during creeping. 
According to Fig. \ref{K vs L}(a), their relationship changes  as the raft prepared by slow pull transits from region I to II. If each ant behaves like a spring, we expect $K\propto W/L$ since the width $W$ of raft measures the number in parallel, while $L$ in series. Therefore, $K\propto 1/L$ is expected as $W$ remains constant in  region I. Entering region II, the drastic shift to an increasing $K\propto L$ behavior derives from the fact that the ants are active and alive. We imagine the larger extension and cracks send a desperate signal to the ants that their life is in peril and a desperate effort is required to save the raft from disintegrating. 
Now let's look at the fast pull data. The  $K\propto 1/L$ in Fast Reg. I-I can be understood by the same argument as Slow Reg. I-I. However, the faster rate of deterioration for raft as one entity in Fast Reg. I-II renders an opposite behavior to Slow Reg. I-II. The neck  is not only narrower, but its width also shrinks faster, as demonstrated by Fig. \ref{K vs L}(b,c). This creation of a weak link in the network overwhelms the last-resort effort by each ant to strengthen its spring constant, and results in a decreasing $K$ with $L$, rather than increasing as Slow Reg. II-II. 

\section{Conclusion and Discussions}

By showing that local ants and dead fire ants do not aggregate on water, we conclude that the ascription of fire ant rafts to Cheerios effect is unsatisfactory. Establishing the fact the raft formation is also possible when fire ants are subject to vertical or horizontal shaking, we tend to buy into the speculation by  E. O. Wilson that such an attractive behavior is triggered by some special pheromone when fire ants sense the danger.

Two key ingredients distinguish the ant raft from membranes composed of normal material. The first is activity: The fire ants are constantly moving, so their set of nearest neighbors is constantly changing. The second is the urge to mend and maintain the integrity of the raft. Insects take time to react to external stimulus and what their neighbors are doing, which sets an intrinsic response time. As a result,
we found in the force-displacement experiment  that
(1) the stress-strain relationship  at low pull speeds is similar to that of  most ductile material, i.e., consisting of elastic, plastic and rupture regions, 
(2) plastic region shortens and the raft  becomes more brittle-like as the pull speed increases,
(3) in analogy to the effect of electric field for liquid crystal, the number of ants whose orientation is aligned to the pull force is  positively correlated with the magnitude of pull force,
(4) the force chains can be mapped for regions II an III.

Furthermore, we found in creep experiment  that
(5) ant raft exhibits zero Poisson ratio, which is beneficial in the engineering of artificial cartilage, ligament and corneal, for region-I  samples, irrespective of their pull speed. Without relying on specific geometry structures, the ants manage such a feat by relocating their idle reserve from the top to bottom layer to prevent the raft from disintegrating. In other words, the zero Poisson ratio is made possible by the fact that fire ants are alive and have the instinct to maintain the integrity of raft.
(6) the effective Young's modulus $\propto 1/L$ can be understood by the spring-network model in region-I when the width of raft is insensitive to extra stress. 
Biomimicry is the emulation of the models, systems, and elements of nature for the purpose of solving complex human problems. Our study of fire ant raft as an active self-heal membrane reveals several interesting mechanical properties, such as zero Poisson's ratio as a purely active effect and the effective Young's modulus $\propto 1/L$ for $L$ is the length of ant raft.


We are grateful to J. Wang, H. Ko, Kh$\rm{\acute{a}}$-$\rm{\hat{I}}$ T$\rm{\hat{o}}$ and J. T. Shy for useful discussions, and Department of Economic Development of Hsinchu City Government for supplying the fire ant samples. Financial support from the Ministry of Science and Technology in Taiwan under Grants No. 111-2112-M007- 025 
is acknowledged.

\end{document}